 \documentclass[preprint2]{aastex}

\shorttitle{Field Rotator Induced Flat Field Systematics}
\shortauthors{Moehler et al.}

\begin{document}

\title{Correction of Field Rotator-Induced Flat-Field Systematics - A
  Case Study Using Archived VLT-FORS Data} 


\author{Sabine Moehler, Wolfram Freudling,
Palle M\o ller, Ferdinando Patat, Gero Rupprecht}
\affil{European Southern Observatory, Karl-Schwarzschild-Str. 2,
  D-85748 Garching bei M\"unchen, Germany}
\author{Kieran O'Brien}
\affil{European Southern Observatory, Casilla 19001, Santiago 19, Chile}
\email{smoehler,wfreudli,pmoller,fpatat,grupprec,kobrien@eso.org}




\begin{abstract} 
        
ESO's two {\bf FO}cal {\bf R}educer and low dispersion {\bf
S}pectrographs (FORS) are the primary optical imaging instruments for
the VLT. They are not direct-imaging instruments, as there are several
optical elements in the light path. In particular, both instruments are
attached to a field rotator. Obtaining truly photometric data
with such instruments present a significant challenge.  In this paper,
we investigate in detail twilight flats taken with the FORS
instruments.  We find that a large fraction of the structure seen in
these flatfields rotates with the field rotator.

We discuss in detail the methods we use to determine the cause of this effect.
The effect was tracked down to be caused by the Linear Atmospheric Dispersion
Corrector (LADC). The results are thus of special interest for designers of
instruments with LADCs and developers of calibration plans and pipelines for
such instruments. The methods described here to find and
correct it, however, are of interest also for other instruments using
a field rotator.

If not properly corrected, this structure in the flatfield may degrade
the photometric accuracy of imaging observations taken with the FORS
instruments by adding a systematic error of up to 4\% for broad band
filters. We discuss several strategies to obtain photometric images in
the presence of rotating flatfield pattern.
\end{abstract}


\section{Introduction}

Systematic differences between flat field images, i.e. high count
level CCD exposures of a smooth/diffuse source, and the actual system
efficiency at any position of the CCD ultimately set a limit for the
photometric accuracy of an imaging instrument unless they are
accurately determined and corrected for. This correction process is
known as ``illumination correction'' and is especially important for
focal reducer type instruments due to their numerous internal reflections
that redistribute diffuse light, an effect sometimes referred to as
``sky concentration'' \citep[see for instance][]{andersen95,koch03}.

The two optical focal reducer instruments FORS1 and FORS2 have been in
operation at the ESO VLT since April 1, 1999, and April 1, 2000,
respectively. During this time a large database of calibration data
has been accumulated. In an effort to improve the overall photometric
accuracy that we offer to ESO users, we are in the process of defining
a procedure to determine and correct for systematics such as sky
concentration \citep{moller,freudling1,freudling2}.

In the course of this work we have found an effect that has
previously not been described in the literature. The effect is seen as
twilight image features that rotate along with the field rotator
position and therefore clearly are created inside the
telescope/instrument system. The effect might not be limited to the
two FORS instruments, but affect other instruments mounted on
telescopes with field rotators. Due to our vast database covering 
about one decade of systematically obtained and documented
calibration data we are able to investigate and quantify the features.

The paper is organized in the following way. In Sec.~\ref{sec:obs}, we describe
the selection and processing of the data used in this investigation. In
Sec.~\ref{sec:res}, we describe how we isolated the rotating part of the flat
fields and investigate the properties  and origin of this structure.
Finally, in Sec.~\ref{sec:photometry}, we discuss the impact of our finding on
photometry and strategies for dealing with this effect.

\section{Data}\label{sec:obs}

The current calibration plan for the FORS instruments specifies that
twilight flat fields have to be taken within seven nights of a science
observation with a given setup. Usually about 4--6 frames are taken,
mostly during evening twilight, but sometimes also during morning
twilight. Hereafter, we refer to these individual images as ``twilight
flats''. In order to eliminate the contributions of field stars on the
jittered sequence of flat fields, the frames are combined using a
median rather than a simple average. Before combining the individual
frames, the master bias is subtracted and each frame is
normalized with the median of its flux. The resulting masterflats are
used for the pipeline processing of the images and also by most users
of FORS data.

We used the individual twilight flats to investigate the properties of
flatfields taken with the FORS instruments.  For that purpose, we
retrieved data from the ESO archive observed between April 1, 1999 and
April 11, 2008 for FORS1 and between May 1, 2000 and November 1, 2008
for FORS2. Within these time intervals, the FORS instruments and the
telescopes they were attached to went through several maintenance
intervals and/or upgrades, and the instruments were moved to different
Unit Telescopes (UTs) of the VLT.  In order to assess the stability of
the twilight flats, we took care to combine only twilight flats taken
between such interventions. The periods we considered are detailed in
Table~\ref{tab:UT}. In addition to the archive data from the
calibration plan, we obtained a specifically designed sequence of
twilight flats for FORS1 with the OII+44 filter between August 31 and September
4, 2007. This filter is an interference filter centered at
372\,nm. These observations are the only ones that were explicitly
taken to analyze the rotating feature and the rotator angle was
changed in roughly 10$^\circ$ steps.

Before investigating the flatfields in more detail, we prepared them
in the following way.  First, each individual image was bias-corrected
using the pre-scan region only and normalized with the median of its
flux (hereafter flux-normalized twilight flat).  The single CCDs of
the FORS instruments were read out via four ports in standard imaging
mode, that yields different gains for the four quadrants (see the
FORS manual\footnote{http://www.eso.org/sci/facilities/paranal/instruments/fors/doc/} for
more details). In these cases the median was determined from the
region from 100,100 (lower left corner) to 900,900 (upper right
corner), covering the first quadrant of the CCD. For the CCD mosaics
we first combined the 2 frames corresponding to one exposure and
extracted the illuminated part.  Within the illuminated part the flux
was determined for the region 500,900 (lower left corner) to 800,1500
(upper right corner). 

\begin{figure}
\plotone{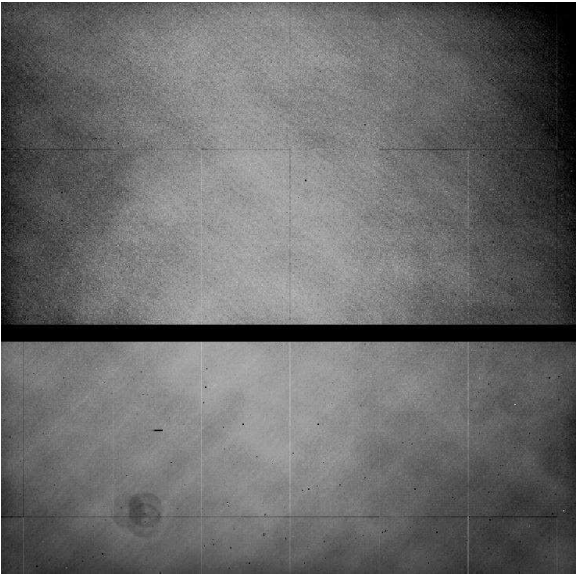}
\caption{The median of the flux-normalized FORS1 $R$ twilight flat
fields at UT2 during the time range from 2007-04-01 to 2007-09-24 with
the two CCDs corrected to the same flux level at 50\arcsec\ distance
from the center of the field. The cuts are at 0.95 and
1.05.\label{fig:F1_UT2_R_2CCDnbw}}
\end{figure}

\begin{figure}
\plotone{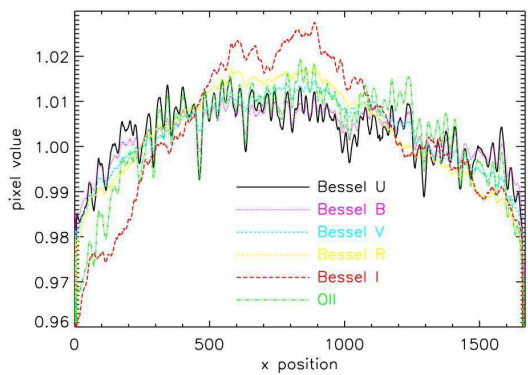}
\caption{Horizontal cuts across the flux-normalized median twilight
flats taken with FORS1 from 2007-04-01 to 2007-09-24.
\label{fig:cuts}}

\end{figure}

A cursory inspection of the flux-normalized flatfields shows that
individual flats differ by as much as 5\% in amplitude. At the same
time, there is a stable component that is similar for all twilight
flats for a given time range between interventions. This stable
component emerges when the median of those twilight flats is computed
and is often dominated by sky light concentration (see
Fig.~\ref{fig:F1_UT2_R_2CCDnbw}). In Fig.~\ref{fig:cuts}, the shape of
the pattern is shown for the broad band and the narrow band filter we
used.  The figure shows cuts through the medians for the Bessell U, B,
V, R and I filters and the OII narrow band filter. The overall shape
and amplitude is similar for all filters, and the total amplitude is
on the order of 3 to 5\%.

In order to investigate the changing component of the twilight flats,
we divided each flat by the median flux-normalized twilight flat for
the corresponding filter (hereafter flatfielded twilight flat).  This
removes any structure of the flatfield which is stable and fixed
relative to the detector.

\section{Changing Component of Twilight Flats}\label{sec:res}

\subsection{Rotating Pattern}\label{sec:pattern}

 Inspection of the flatfielded twilight flats revealed that some of
the remaining structures seem to change their position between
flatfields taken in the same night. This observation immediately rules
out that these features are caused by spatial sensitivity variations
on the detector, or vignetting on structures which are physically
fixed relative to the detector. Instead, at least some of the
variations in the flatfields must be caused by moving parts within the
telescope or FORS instrument. The alt/az mounting of the VLT UT
telescopes causes field rotation, which is compensated by rotating the
instrument accordingly. Fixed components along the light path
(e.g. the Linear Atmospheric Dispersion Corrector or the M3) will
therefore rotate with respect to the detector of the instrument. To
test whether the flatfield structures are related to the field
rotation, we compared twilight flats taken with different
rotator angles.

\begin{figure}
\plotone{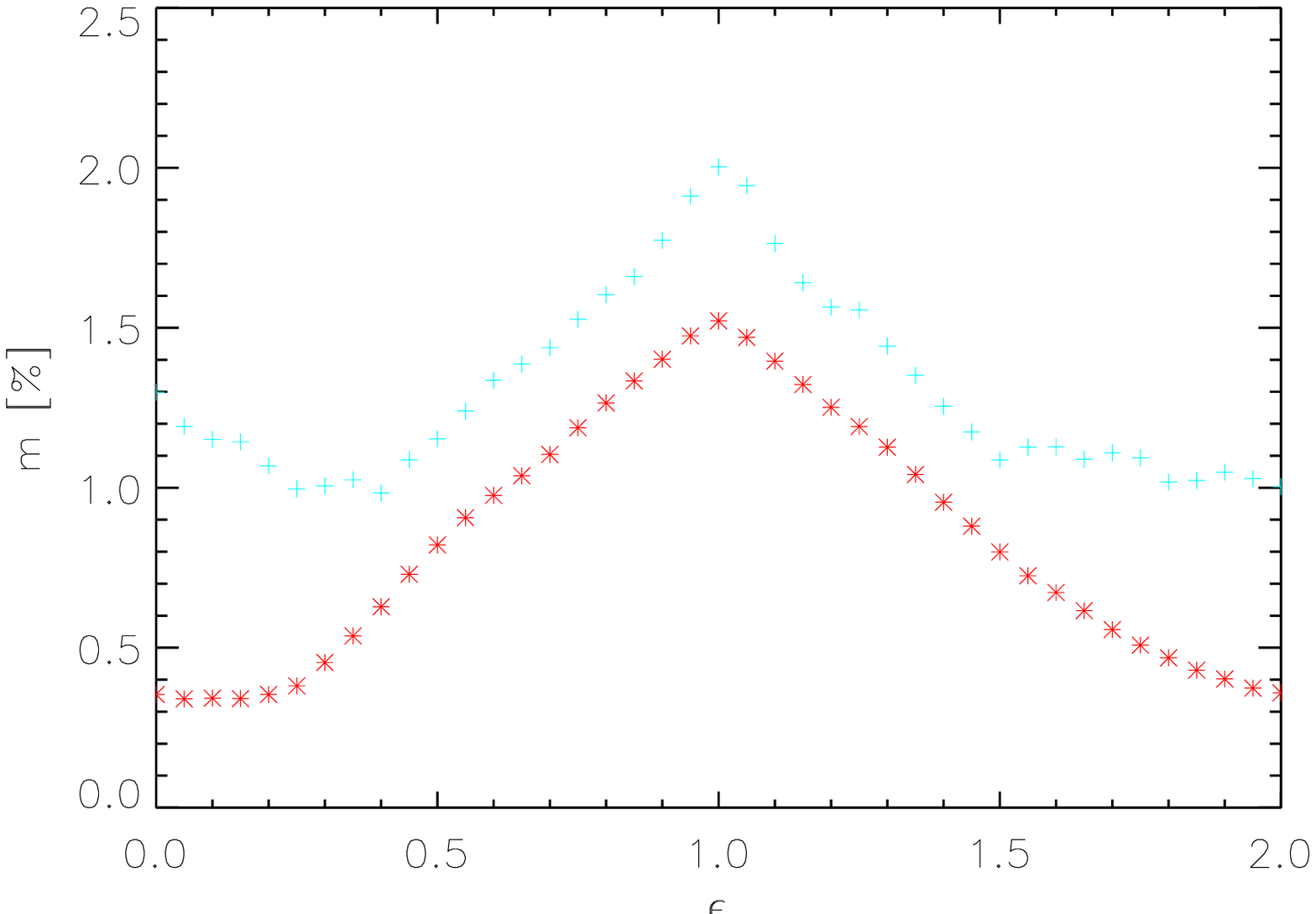}
\plotone{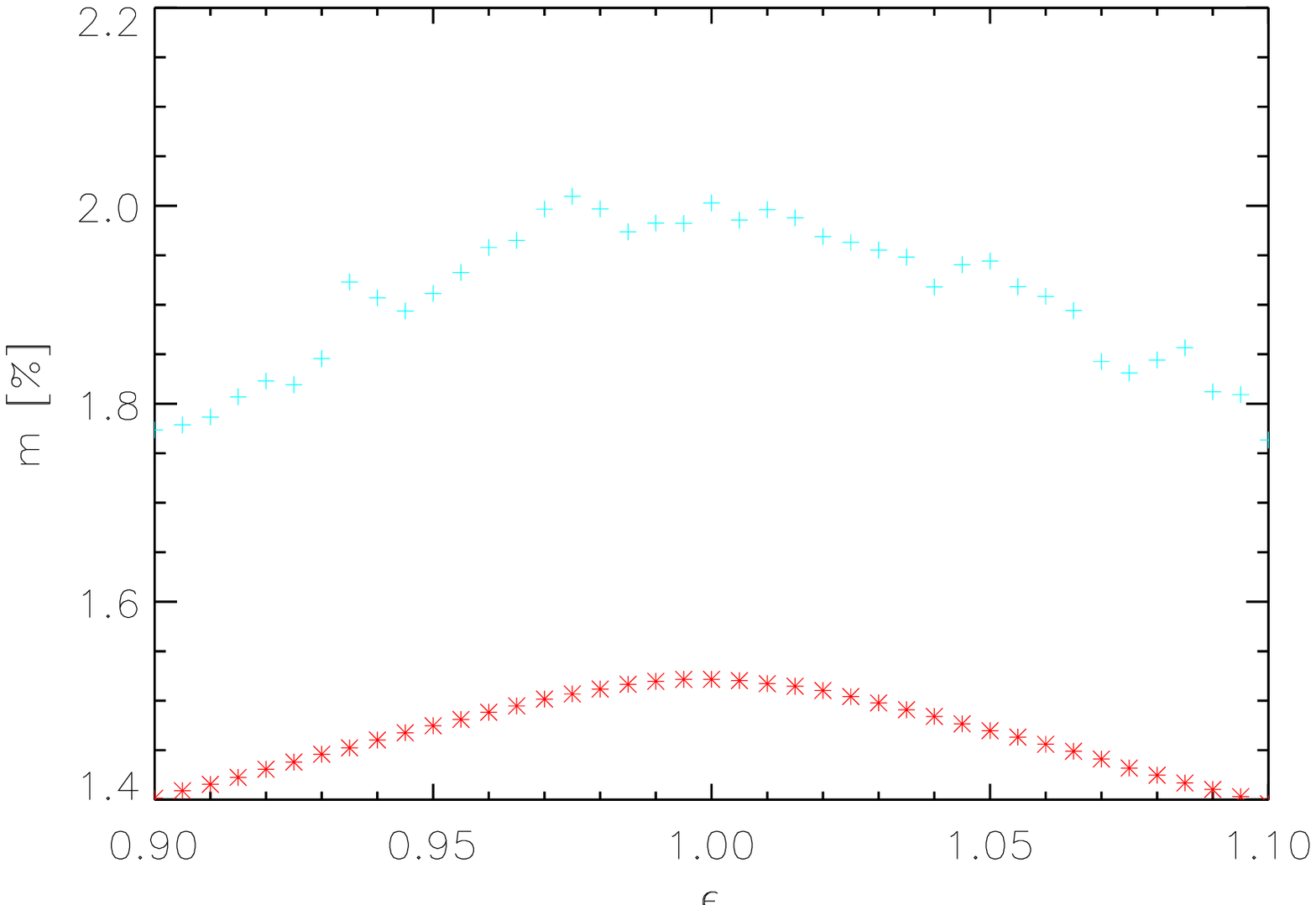}
\caption{Amplitudes measured in the median of rotated flatfielded
twilight flats as a function of rotation factor. Crosses ($+$) are the
differences between maximum and minimum pixel values for each
rotation angle, whereas the stars are the amplitudes of structures
estimated as the difference between the 99$^{\rm th}$ and 1$^{\rm st}$
percentiles. A rotation factor of unity indicates that each flatfield
has been rotated by the negative of the rotator angle. The top panel
show the whole range of values for $\epsilon$, while the lower panel
zooms in on the peak of the distribution. For this test we used the
data set FORS1@UT1, 2000-08-01 to 2001-02-25.
\label{fig:amps}}
\end{figure}

In order to investigate the exact relation between the orientation of
structures in the twilight flats and that of the field rotator, we took a set
of twilight images with a maximum of 5 images per 5$^\circ$ interval in rotator angle.
These were then counter rotated with the angle of the field rotator multiplied
with a factor $\epsilon$ between 0 and 2. Then we computed the median of these
counter rotated flatfielded twilight flats and measured the amplitude of the
structures as the difference between the 99$^{\rm th}$ and 1$^{\rm st}$
percentile. Only pixels within a centered circle with a diameter equal to the
dimension of the image were used for that purpose.  The amplitudes as a
function of the rotation factor $\epsilon$  are shown in
Fig.~\ref{fig:amps}. It can be seen that the amplitudes of structures peak when
the de-rotation angle is exactly identical to the rotator angle. We therefore
use rotations with the opposite of the exact rotator angle to
isolate the rotating pattern (hereafter RP).

\begin{figure}
\plotone{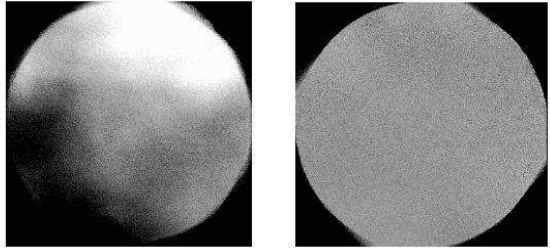}
\caption{{\bf Left:} The median of the flatfielded FORS1 $B$
twilight flat fields rotated to rotator angle of 0 at UT1 during the
time range from 2000-08-01 to 2001-02-25. {\bf Right:} Same as in left
panel, but with the individual images rotated to random rotator
angles. Both images are displayed with cuts of 0.995 and 1.005.
\label{fig:F1_B_UT1_B_rotator}}
\end{figure}

We applied this procedure to the data sets described in
Table~\ref{tab:UT}.  We compared the structures in the resulting image
with the median of the same frames, but rotated by a random rotation
angle.  One example of such a comparison is shown in
Fig.~\ref{fig:F1_B_UT1_B_rotator}. The amplitude of structures in the
median of the counter rotated flatfielded frames is typically between
about 0.6\% and 2.0\%, whereas no structure can be recognized in the
median of randomly rotated flatfielded frames.

\begin{figure}
\plotone{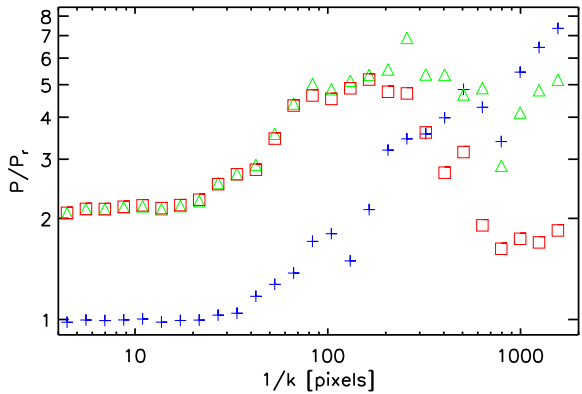} 

\caption{Normalized power spectra as a function of scale length
    derived from a flatfielded twilight flat field, that has been
    corrected for the stable component (same data set as for
    Fig.~\ref{fig:amps}).  The crosses ($+$) represent the power
    spectrum of the RP, the triangles that of the average of twilight
    flats with the same rotator angle, and the squares of the average
    with the RP removed (see text). All power spectra were computed
    within the central circle on the images, and divided by the power
    spectrum of the randomly rotated average.
\label{fig:powerspectra}} 

\end{figure}

So far, we have identified two components of the flatfields, one is the pattern
fixed relative to the detector, and the other one is the RP. An interesting
question is on what scales the RP becomes relevant, and whether all
structure in the flatfield which is not fixed to the detector can be attributed
to the RP. To investigate this, we computed the power spectrum within the
central circle of the isolated RP, and divided it by the power spectrum of the
same region in the randomly rotated average.  The result is shown as crosses
($+$) in Fig.~\ref{fig:powerspectra}. It can be seen that the range of scales
present in the RP is from about 30 pixels to the size of the CCD. This
power spectrum should be compared to that in a flatfield with a single rotator
angle.   For that purpose, the power spectrum of twilight flats with rotator
angles between 135$^\circ$ and 145$^\circ$, is shown as triangles  in
Fig.~\ref{fig:powerspectra}.  A comparison of these two power spectra reveals
that the structure in the flatfields at scales above about 300 pixels is
dominated by the RP. At intermediate scales between~30 and~300 pixels, the
amplitude of the RP is small compared to other structure in the flatfield. To
illustrate this more clearly, we removed the RP from the individual flatfielded
flats with rotator angles between 135$^\circ$ and 145$^\circ$ after rotating it
in place, and then computed the average. The power spectra of the RP corrected
average flatfielded flat are shown as squares in Fig.~\ref{fig:powerspectra}.
It can be seen that most of the power at scales larger than 300 pixels can be
removed when correcting the isolated RP at the correct rotator angle.

\subsection{Rotating Pattern for different filters}\label{sec:filters}

\begin{figure}
\plotone{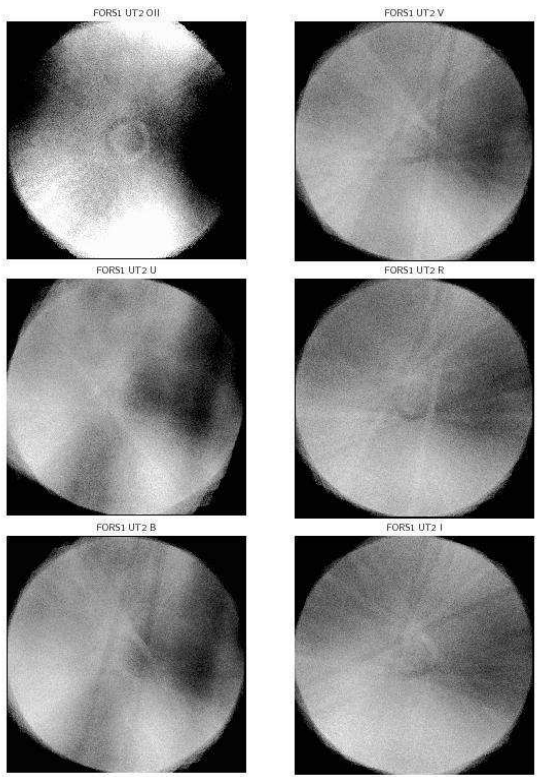} 
\caption{Comparison of flatfielded twilight flat fields rotated to rotator
angle~0 for the broad-band filters UBVRI and the narrow-band filter OII
observed with FORS1 at UT2 from 2007-04-01 to 2007-09-24. All images are
displayed with cuts of 0.995 and 1.005. See text for further
  details.\label{fig:fors1}} 
\end{figure}

\begin{figure} 
\plotone{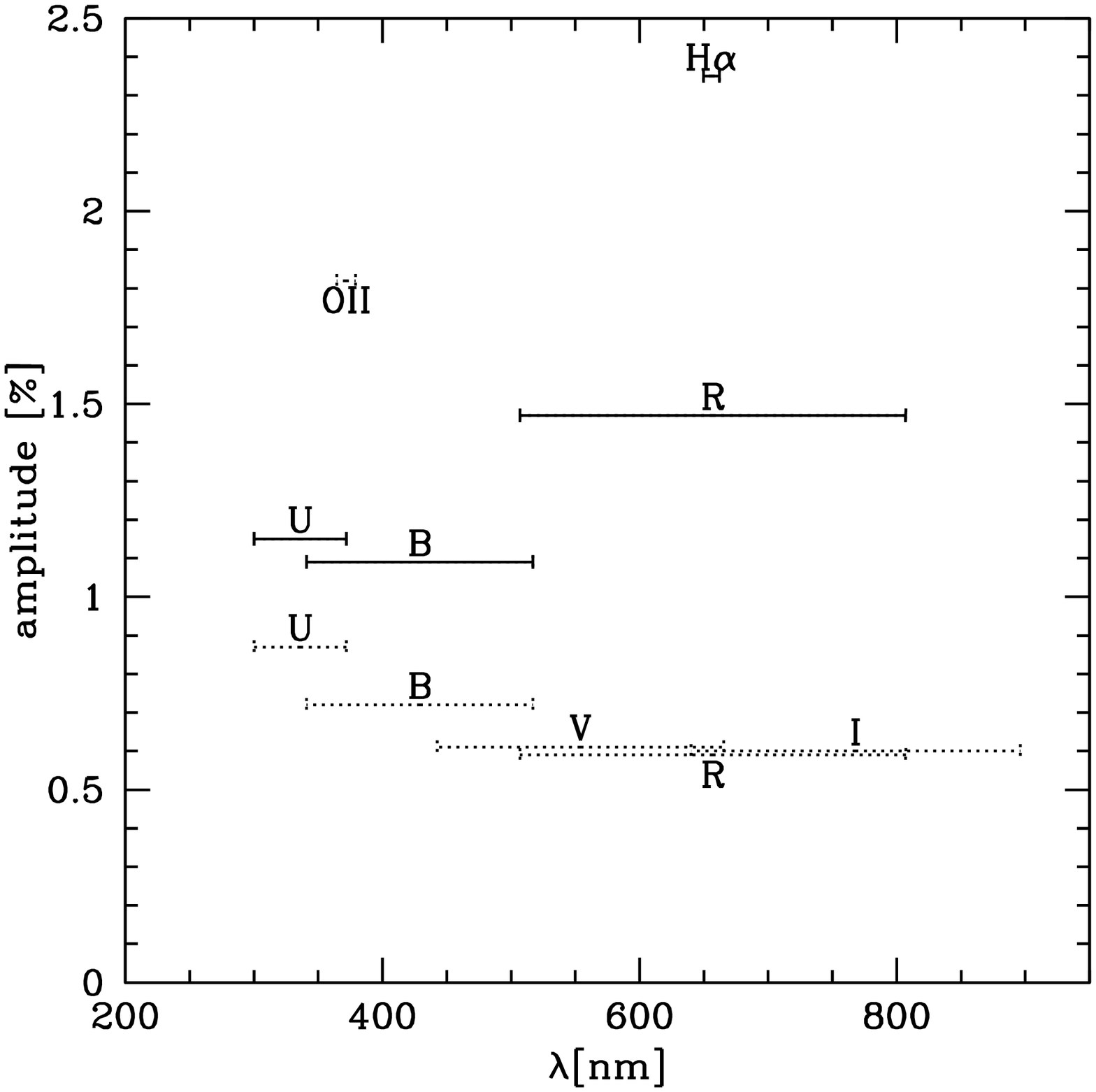} 
\caption{Amplitudes of the rotating pattern as a function of the
wavelength range of the filters for the time ranges listed in
Tab.~\ref{tab:filter}. The FWHM filter width is shown for each filter,
FORS1/2 data are marked by dotted lines and solid lines, respectively.
\label{fig:filters}}
\end{figure}

The next question we wish to address is whether the shape and
amplitude of the RP depend on the filter. We therefore isolated the RP
with the method described in Sect~\ref{sec:pattern} for all filters
within the periods that include a sufficient number of narrow band
observations.  In Fig.~\ref{fig:fors1}, we show the RPs for one of
these periods.  It is clearly seen that the amplitude as well as the
shape of the structure vary smoothly with wavelength for the broad
band filters.  The amplitudes for different filters are listed
Tab.~\ref{tab:filter}, and the variation in amplitude are illustrated
in Fig.~\ref{fig:filters}.  The amplitudes are about 1.8\% for the
narrow-band filter, and below 1.5\% for the broadband filters. This
trend has to be considered when searching for the cause of the RP (see
Sec.~\ref{sec:residuals}).

This larger amplitude of the RP seen in the narrow band filters means
that more photons were detected in the RP relative to photons detected
in the underlying twilight image.  One possible way to explain this is
that while the underlying image only detects photons within the
passbands of the narrow band filters, the RP is made up of photons
that do not pass through the filter and therefore has the full
bandwidth of unfiltered twilight.  This explanation implies a
correlation of the amplitude with the fraction of photons that pass
through the filter. Such an effect has been reported by \cite{fynbo}
for narrow band observations at the NOT telescope.  This fraction of
sky photons that pass through a filter can be computed from the width
of the filter bandpass, the mean transmission of the filter, the mean
CCD efficiency over the bandpass, the relative brightness of the
twilight sky within the bandpass and the total sky brightness and
exposure time.  To estimate this number, we used the twilight sky
brightness at different bandpasses from \cite{patat}, and the filter
and CCD characteristics for FORS, which are available on the World
Wide
Web\footnote{http://www.eso.org/sci/facilities/paranal/instruments/fors/inst/}.

We then computed the ratio $r$ of the number of photons available outside the
filter to the number of photons detected after passing through the filter as

\begin{equation}
r={\frac{\int N_{p}(\lambda)f(\lambda) {\rm d} \lambda}{\int{N_p(\lambda) {\rm d}\lambda}}}
\end{equation}

where $N_p(\lambda)$ is the detected photon rate as a function of wavelength,
and $f(\lambda)$ is the filter throughput curve. The photon rate was computed
from 

\begin{equation}
N_{p}(\lambda)= S(\lambda)\cdot R_{\rm CCD}(\lambda)
\end{equation}

where $S(\lambda)$ is the twilight sky brightness expressed in photons per
wavelength, and $ R_{\rm CCD}(\lambda)$ is the response of the CCD.

\begin{figure} 
\plotone{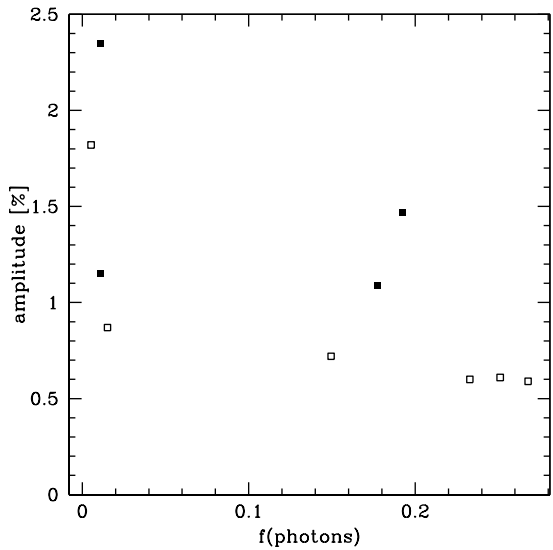} 

\caption{Amplitudes of rotating pattern as a function of the fraction
of sky photons that pass through the filter. Different symbols are
used for the two periods listed in Tab.~\ref{tab:filter},
FORS1/2 data are distinguished by open and filled symbols,
respectively.
\label{fig:amps_photons}} 
\end{figure}

In Fig.~\ref{fig:amps_photons}, we plot the measured amplitude of the
RP versus the fraction of sky photons that pass through the filter.
It can be seen that there is some trend in the sense that a when only
a small fraction of the photons pass through the filter, the
amplitudes tend to be higher.  However, the large scatter make this
test inclusive. In Sec.~\ref{sec:photometry}, we will discuss a more
sensitive test to determine whether the RP is likely to be caused by
scattering.

\subsection{Stability of Rotating Pattern}

\begin{figure}
\plotone{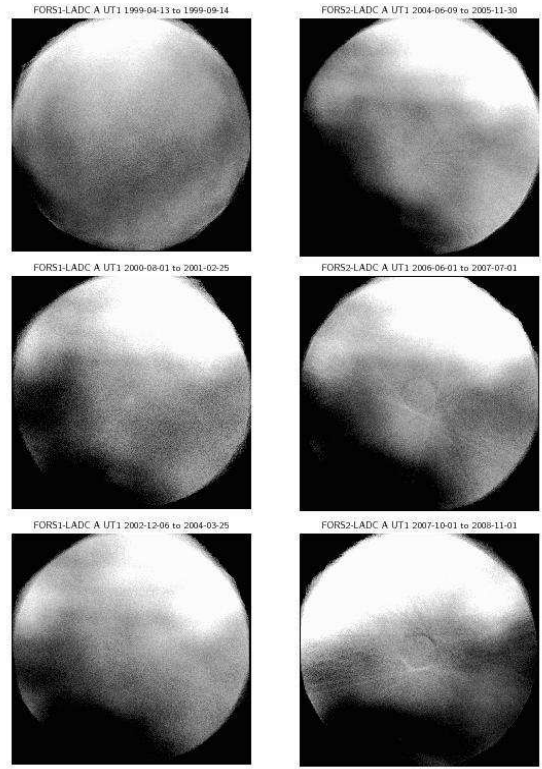}

\caption{Comparison of flatfielded B\_BESS twilight flat fields
rotated to rotator angle 0 observed at UT1. The data on the left are
from FORS1 (1999-04-13~to~1999-09-14, 2000-08-01~to~2001-02-25,
and 2002-12-06~to~2004-03-25 from top to bottom). The data on the
right are from FORS2 (2004-06-09~to~2005-11-30,
2006-06-01~to~2007-07-01, 2007-10-01~to~2008-11-01 from top to
bottom). All images are displayed with cuts of 0.995 and
1.005.\label{fig:ut1}}
\end{figure}

\begin{figure}
\plotone{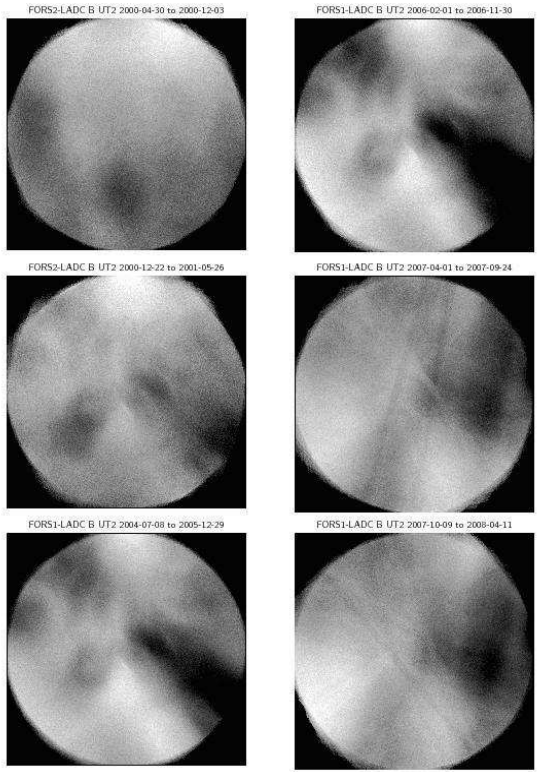}

\caption{Same as in Fig.~\ref{fig:ut1}, but for UT2. The data on the
  left, top and center, are from FORS2 (2000-04-30~to~2000-12-03,
  2000-12-22~to~2001-05-26 from top) The data on the bottom left and
  on the right are from FORS1 (bottom left: 2004-09-01~to~2005-12-29;
  right, top to bottom: 2006-02-01~to~2006-11-30, 2007-04-01~to~2007-09-24,
  2007-10-09~to~2008-04-11).\label{fig:ut2}}
\end{figure}

\begin{figure}
\epsscale{0.45}
\plotone{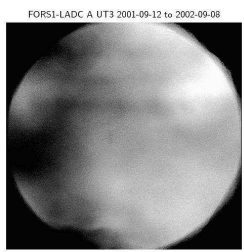} 
\caption{Same as in Fig.~\ref{fig:ut1}, but for UT3 (FORS1).  The date range 
 is 2001-09-12~to~2002-09-08.
\label{fig:ut3}}
\end{figure}

\begin{figure}
\epsscale{1.}
\plotone{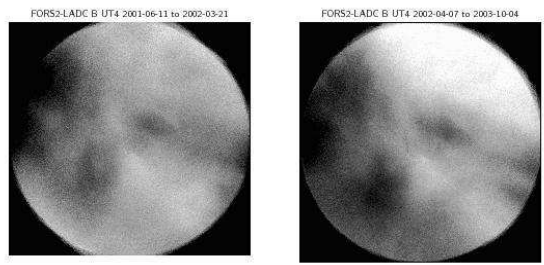} 
\caption{Same as in Fig.~\ref{fig:ut1}, but for UT4 (FORS2).  The date
ranges are 2001-06-11~to~2002-03-21 (left) and
2002-04-07~to~2003-10-04 (right).\label{fig:ut4}}
\end{figure}

The finding that some of the structures in twilight flats rotate with
the rotator angle has significant impact on the photometry with either
of the FORS instruments. To correct for this effect, it is important
to know how stable this pattern is with time. Both instruments have
been moved between the telescopes which comprise the four unit VLT.
Both FORSs use Linear Atmospheric Dispersion Correctors
\citep[LADCs]{avila97} which are fixed to the telescope and mounted in
front of the instrument. The two existing LADCs, called LADC-A and
LADC-B, have also been switched
between the two FORSs once (June 2004).  The different
combinations of LADCs, telescopes and instruments can be used to
search for correlations between the RP and the use of those optical
components.  We used the B Bessell filter for this
investigation. While the amplitude of the signal is stronger in the U
Bessell filter, there are many more twilight flats observed for the B
Bessell, so we can achieve a more homogeneous angle distribution and a
better signal-to-noise ratio.  In Figures~\ref{fig:ut1} --~\ref{fig:ut4}, the
rotating structures are shown for different periods when the FORS
instruments were mounted at the VLT units UT1 to UT4. The amplitudes of
the structures are listed in Tab.~\ref{tab:rot}, together with the
combination of instrument, UT and LADC used in each case.

Inspection of the figures shows that the general structures frequently
but not always change when any changes were made to the instrument. In
most cases, these changes involve removing the instrument from the
field rotator and remounting it. Slight changes in the optical
alignment might therefore explain these differences. The overall shape
of the RP seems to be more strongly correlated with the LADC than with
either the UT or the instrument used.  Structures observed with FORS1
and FORS2 look similar when the LADC A was used at UT2, UT3 or UT4.
At UT1 with LADC A (Fig.\ref{fig:ut1}), both FORSs show a gradient of
increasing flux from the lower left corner to the upper right corner
for data observed after 1999. The amplitude of this effect increased
when the LADC was switched from FORS1 to FORS2
(cf. Table~\ref{tab:rot}).  On the other hand, for UT2 and UT4 in
combination with LADC B (Figs.~\ref{fig:ut2} and~\ref{fig:ut4}) one
can see the same features, a slanted ``1'', in FORS1 data observed
between July 2004 and December 2006 and in FORS2 data observed between
December 2000 and May 2001 (UT2) and between June 2001 and October
2003 (UT4). A similar feature can be observed in the newest FORS1 data
(since April 2007), although the ``1'' appears to be flipped and
rotated.  The feature is not present in the earliest FORS2 data at
UT2, which look more similar to the FORS1 data from UT3
(Fig.~\ref{fig:ut3}) and show a slope similar to the UT1 data, albeit
with some variation in the low flux region.

\subsection{Contamination of the LADC}

\begin{figure} 
\epsscale{1.} 
\plotone{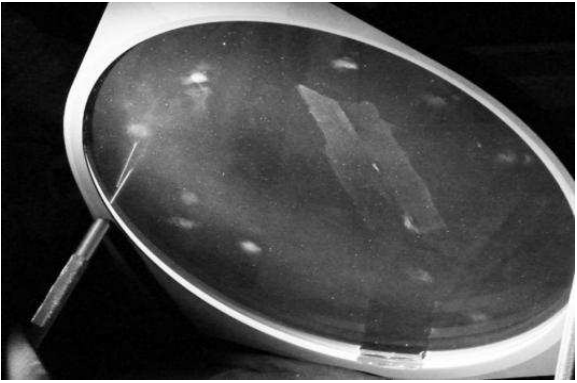} 
\caption{The lower prism of LADC B illuminated by a flashlight, photograph
  taken on 2009 Jun 29.
\label{fig:LADC}} \end{figure}

Under normal circumstances, the LADC is not accessible unless the FORS
instrument is dismounted. After we identified the LADC as a possible
cause for the RP, we visually inspected LADC B, which has been
de-commissioned with FORS1 on March 31, 2009. Fig.~\ref{fig:LADC}
shows a picture of the lower prism illuminated with a flash
light. Comparing the smudges seen in this picture to the structures
visible in Figs.~\ref{fig:fors1}, \ref{fig:ut2}, and \ref{fig:ut4}
suggests that the LADC is responsible for a major part of these
structures.

\section{Implications for Photometry}\label{sec:photometry}

\subsection{RP and Photometric Zeropoint Variations}\label{sec:residuals}

The observed RP raises the question how accurate relative
photometry can be obtained from FORS images. The obvious question is
whether the RP is a faithful representation of throughput variations
across the detector, or whether it is an additive defect in the
flatfields that should be removed.  To address this issues, we
analyzed a set of dithered R-band observations of Stetson standard
fields \citep{freudling1}. Correlating the measured magnitudes of
stars with features of the RP can determine whether the RP is additive
or multiplicative, and thereby decide on the appropriate correction
procedure.

The details of data taking and reduction are given in \citet{freudling1,
freudling2,freudling3}. Here, we only give a brief summary.
In total, 30 images of the field were taken on a 5$\times$5 grid
covering a total area of about 12\arcmin$\times$12\arcmin, including
several rotations of the field.  We removed any structure larger than 100
pixels from the masterflat, and the resulting masterflat was used to
flatfield all the images.  We then measured instrumental aperture magnitudes
for any star that is included on at least two of the images.  In total, there
are about 10\,000 measured magnitudes of 900 unique stars.  The measured
magnitudes were then used to fit a model of the illumination pattern, relative
magnitude zero points of the stars and a zero point for each image.  The model
for the illumination pattern was a two-dimensional third order polynomial. The
scales which can be fitted with such a model are too large to remove any
possible sensitivity variations on the scales of the RP. We computed the
residuals from the fit $\Delta m= m-m_{\rm st} - m_{\rm i} - i(x,y)$, where $m$
are the measured instrumental magnitudes, $m_{\rm st}$ are the relative
magnitudes of the stars, $m_{\rm i}$ the relative zero points of the images,
and $i(x,y)$ is the model of the illumination pattern.

\begin{figure} 
\epsscale{1.} 
\plotone{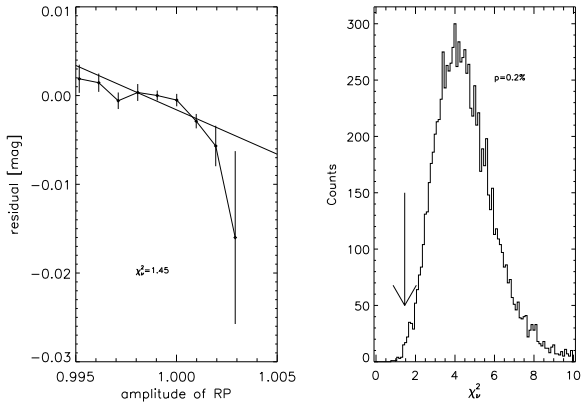} 
\caption{Left panel: Magnitude residuals as a function of pixel value of
the RP.
The data points are the averaged magnitude residuals for a given level
of the RP, error bars
are the 1$\sigma$ uncertainties of the mean.  The solid line is the 
fit of a line with a slope of -1 to the data points. The reduced $\chi^2$ 
of the fit is 1.45.
 Right panel: Distribution of reduced $\chi^2$ for 10\,000 fits to
 residuals versus pixel values of the RP. The positions of the stars
 were randomly assigned for each each of the fits. The arrow marks
 the $\chi^2_\nu$ of the original data as shown in the left panel.
\label{fig:residuals}} 
\end{figure}

We then plotted these residuals with the corresponding pixel value of the
normalized flatfields taken with the same rotator angle as the stellar images.
The results are shown in left panel of Fig.~\ref{fig:residuals}.  It can be
seen that there is a correlation between the two quantities in the sense
that measured magnitudes of the same stars are brighter when they are in areas
where the RP is bright, and fainter where the RP is faint. A line with a slope
of $-1$ in Fig.~\ref{fig:residuals} shows the expected relation if the RP 
directly represents sensitivity variations across the detector.  The reduced
$\chi^2_\nu$ from this line is 1.45. This suggests that the RP is indeed a
valid part of the flatfield, i.e. the pattern is multiplicative and is directly
related to the photometry.

In order to evaluate the significance of this result, we repeated the
described procedure with the positions of the stars on the detector
randomly drawn. In this case, we do not expect any relation between
the residuals and the values on the RP. Residuals from fitting a line
with a slope $-$1 therefore will be large, and this can be detected by
a large values for $\chi^2_\nu$.  For each realization, we computed
the $\chi^2_\nu$ of the fit in the same way as we did with the
original data. The distribution of $\chi^2_\nu$ for 10\,000
realizations is show in the right panel of
Fig.~\ref{fig:residuals}. Only in about 0.2\% of all realization was
the $\chi^2_\nu$ as low as for the original data. As an additional
test, we repeated the procedure with the original data, but the
RP rotated by 90$^\circ$. Again, there was no correlation between the
residuals and the RP. We therefore conclude that at least for the
R\_BESS filter we tested, the RP presents a sensitivity
variation across the detectors.

However, as discussed in Sec.~\ref{sec:filters} we also see evidence
for a scattered component which shows up relatively strongly for
narrow band filters.  Therefore, the RP for the narrow band filters
might include both multiplicative and additive components.

\subsection{Data Reduction Strategy}\label{sec:strategy}

One consequence of the findings discussed here is that the RP will
significantly affect the master flats unless the rotator angles are
carefully controlled.  Applying a flatfield that includes the RP to
imaging data with a different rotator angle will introduce shifts of
the photometric zeropoint with amplitudes twice as large as the RP,
i.e. up to almost 4\% across the inner circle on the detector for
broad band filters.  There are several conceivable strategies to
address this problem.  The first is to try to isolate the RP as we
have described, rotate it to the rotator angle of each individual
twilight flat, and correct the master flats.  The disadvantage of this
approach is that it is not applicable to the corners of the detectors
where the RP cannot be isolated. The second strategy is to remove any
structure in the twilight flats on the scale of the structures seen in
the RP (see Fig.~\ref{fig:powerspectra}), and then determine the
larger scale illumination correction through independent
observations. The disadvantage of this approach is that it is
difficult to determine the illumination correction with sufficient
spatial resolution.  Finally, one could address the RP matching the
orientation of flat fields to the corresponding science images. For
reasons of efficiency, the implementation of such a strategy requires one
to restrict the orientation of science data to the small number of
selected rotator angles, for which flatfields are available.  We
consider this solution to be the best strategy.

We also recommend this strategy of searching for twilight flats with
  rotator angles close to those of the science data to users, who
  already have FORS data or retrieve them from the archive. They must,
  however, avoid mixing data across interventions. If there are
  no flat fields in the given time range with a rotator angle similar
  to that of the science data, the next best solution is to construct
  a master twilight flat by combining twilight observations with a
  large range of rotator angles to smear out the RP.

\section{Summary and Conclusions}\label{sec:conclusions}

Using archived calibration data from more than eight years we
have shown that the master twilight flats regularly produced from FORS
observations include structures that are not related to sensitivity
variations across the detector. Part of this is caused by field
illumination effects, which can be corrected using standard field
illumination correction \citep[for details see][]{freudling1} but in
addition we find a pattern which rotates with the setting of the field
rotator. This pattern is stable in the absence of instrument interventions,
but occasionally changes when work on the instrument has been
performed.

Obviously the source of this additional pattern must be either partly
or completely external to the instrument itself in order to be able to
follow the field rotator. Possible candidates were therefore the guide probe
or structures inside or on the M3 tower, e.g.
reflections off some structure related to the M3 mirror or to
the LADC. Alternatively the effect may be caused directly by
transmission variations in the optical surfaces of the LADC itself.
Since the method by which one would correct for the effect is very
different for reflections (additive effect) or transmission variations
(multiplicative), we have deemed it important to determine its source.

Data taken through a narrow band filter suggested at first that
reflections were the main cause, but data taken through a second
narrow band filter did not strongly support this suspicion. There is
evidence that the rotating patterns followed the LADCs when they moved
between the two instruments. Further a direct inspection of one of the
LADCs, which has now been dismounted and decommissioned, showed a
structure on its coating which well resembles the pattern seen in the
flat fields. A last test finally showed that comparison to photometric
stellar data confirmed that the pattern is multiplicative,
i.e. consistent with transmission variations.  Having thus identified
the cause we conclude that if it is left uncorrected in the
worst case it could cause systematic errors of up to 4\% in the photometry.
Future photometric observations with the FORS2 instrument will have to
take this newly discovered feature into account in order to remove or
minimize its impact. Several options for corrective action exist and
are described in this paper in Sect.~\ref{sec:strategy}.

One might well expect that a similar detailed analysis of archived
calibration data from other long-term-stable instrument/telescope
configurations on alt/az mounted telescopes would produce similar
results, in particular if these also include LADCs (e.g. LRIS on
Keck). We recommend that such analysis should be performed on all
instruments used for photometry as part of the general health check
and trending analysis when sufficient calibration data are available. Such
effects could be stronger or weaker on other instruments than were
found here, and they could be both additive and multiplicative as
described above. In this paper we have provided a detailed description
of how such an analysis is best performed, how one may best determine
the source of such effects, and how one may correct for them.

\acknowledgements{Acknowledgments: We thank Hans Dekker, Martino
Romaniello, and Andreas Kaufer for valuable discussions. We
highly appreciate the help of the staff on Paranal in locating and
examining the LADC-B.}













\clearpage
\begin{deluxetable}{llllllr}
\tabletypesize{\scriptsize}
\rotate
\tablecaption{Combinations of telescope and FORS\label{tab:UT}}
\tablewidth{0pt}
\tablehead{
\colhead{Instrument} & \colhead{Unit Telescope} & \colhead{start} &
\colhead{end} & \colhead{time range used} & \colhead{event} &
\colhead{\# frames}}
\startdata
FORS1 & UT1-Antu & 1999-04-01 & 2001-07-31 & 1999-04-13 \ldots 1999-09-14 & 1999-10-25 FORS1 maintenance & 142 \\
 & & & & 2000-08-01 \ldots 2001-02-25 & 2001-03-25 mirror re-coating & 137\\
 & UT3-Melipal & 2001-08-02 & 2002-10-19 & 2001-09-12 \ldots 2002-09-08 & 2002-10-19 move from UT3 to UT1 & 266 \\
 & UT1-Antu & 2002-10-22 & 2004-05-30 & 2002-12-06 \ldots 2004-03-25 & 2004-06-04 move from UT1 to UT2 & 234 \\
 & UT2-Kueyen & 2004-06-06 & 2009-03-31 & 2004-09-01 \ldots 2005-12-29 & 2006-01-27 FORS1 maintenance & 247 \\
 &  &  &  & 2006-02-01 \ldots 2006-11-30 & 2006-12-01 mirror recoating & 201 \\
 &  &  &  & 2007-04-01 \ldots 2007-09-24 & 2007-04-01 new CCD mosaic & 46/106/134/127/129/141 \\
 &  &  &  &  & 2007-09-24 FORS1 maintenance & (OII/U/B/V/R/I)\\
 &  &  &  & 2007-10-09 \ldots 2008-04-11 & 2008-04-11 b\_HIGH replaces B\_BESS as standard filter & 132 \\
 & & & & & & \\
FORS2 & UT2-Kueyen & 2000-03-25 & 2001-06-01 & 2000-03-30 \ldots 2000-12-03 & 2000-12-08 FORS2 maintenance & 155 \\
 & & & & 2000-12-22 \ldots 2001-05-26 & 2001-06-02 move from UT2 to UT1 & 125\\
 & UT4-Yepun & 2001-06-05 & 2004-05-29 & 2001-06-11 \ldots 2002-03-21 & 2002-03-29 new CCD mosaic & 199 \\
 & & & & 2002-04-07 \ldots 2003-10-04 & 2004-05-31 move from UT4 to UT1  & 289 \\
 & UT1-Antu & 2004-06-06 & & 2004-06-09 \ldots 2005-11-30  & 2005-12-15 mirror recoating & 256\\
 &          &            & & 2006-06-01 \ldots 2007-07-01 & 2007-07-02 mirror recoating & 154\\
 &          &            & & 2007-10-01 \ldots 2008-11-01 & 2008-11-06 mirror recoating & 187\\
\enddata
\end{deluxetable}

\clearpage

\begin{deluxetable}{lrrr}
\tabletypesize{\scriptsize}
\tablecaption{Amplitude of the rotator effect with
  wavelength and bandwidth for FORS1@UT2 (2007-04-01 \ldots
  2007-09-24) and FORS2@UT4 (2002-04-07 \ldots 2003-10-04)\label{tab:filter}} 
\tablewidth{0pt}
\tablehead{
\colhead{Filter} & \colhead{Central Wavelength (nm)} &  \colhead{FWHM
  (nm)} & \colhead{amplitude (\%)}\\} 
\startdata
\multicolumn{4}{c}{FORS1}\\[2mm]
\hline\\
U\_BESSELL & 336.0 &  36.0 & 0.87  \\
OII        & 371.7 &   7.3 & 1.82  \\
B\_BESSELL & 429.0 &  88.0 & 0.72  \\
V\_BESSELL & 554.0 & 111.5 & 0.61  \\
R\_BESSELL & 657.0 & 150.0 & 0.60  \\
I\_BESSELL & 768.0 & 138.0 & 0.61  \\
\hline\\[2mm]
\multicolumn{4}{c}{FORS2}\\[2mm]
\hline\\
U\_SPECIAL & 362.0 & 29.0  & 1.15 \\
B\_BESSELL & 429.0 & 88.0  & 1.09\\
R\_SPECIAL & 655.0 & 165.0 & 1.47\\
H$_\alpha$ & 656.3 & 6.1   & 2.35\\
\enddata
\end{deluxetable}

\clearpage
\begin{deluxetable}{lllrlc}
\tabletypesize{\scriptsize}
\tablecaption{Amplitude of the rotator effect with
  telescope-instrument-LADC combination\label{tab:rot}}
\tablewidth{0pt}
\tablehead{
\colhead{Instrument} & \colhead{Unit Telescope} &  \colhead{time range
  used} & \colhead{amplitude (\%)} & \colhead{features} & \colhead{LADC}\\} 
\startdata
FORS1 & UT1-Antu & 1999-04-13 \ldots 1999-09-14 & 1.04 & --- & A \\
FORS1 & UT1-Antu & 2000-08-01 \ldots 2001-02-25 & 1.73 & slope & A \\
FORS1 & UT1-Antu & 2002-12-06 \ldots 2004-03-25 & 1.75 & slope & A \\
FORS2 & UT1-Antu & 2004-06-09 \ldots 2005-11-30 & 1.96 & slope & A \\
FORS2 & UT1-Antu & 2006-06-01 \ldots 2007-07-01 & 2.03 & slope & A \\
FORS2 & UT1-Antu & 2007-10-01 \ldots 2008-11-01 & 1.97 & slope & A \\
 & & & \\
FORS2 & UT2-Kueyen & 2000-04-30 \ldots 2000-12-03 & 0.72 & ``spotty slope'' &B \\
FORS2 & UT2-Kueyen & 2000-12-22 \ldots 2001-05-26 & 0.87 & slanted ``1'' & B\\
FORS1 & UT2-Kueyen & 2004-07-08 \ldots 2005-12-29 & 0.97 & slanted ``1'' & B\\
FORS1 & UT2-Kueyen & 2006-02-01 \ldots 2006-11-30 & 0.72 & slanted ``1'' & B\\
FORS1 & UT2-Kueyen & 2007-04-01 \ldots 2007-09-24 & 1.19 & slanted ``1'', flipped & B\\
FORS1 & UT2-Kueyen & 2007-10-09 \ldots 2008-04-11 & 0.90 & slanted ``1'', flipped & B\\
 & & & \\
FORS1 & UT3-Melipal & 2001-09-12 \ldots 2002-09-08 & 1.04 & ``spotty slope'' & A\\
 & & & \\
FORS2 & UT4-Yepun & 2001-06-11 \ldots 2002-03-21 & 0.87 & slanted ``1'' & B \\
FORS2 & UT4-Yepun & 2002-04-07 \ldots 2003-10-04 & 1.10 & slanted ``1'' & B \\
\enddata
\end{deluxetable}


\begin{thebibliography}{}

\bibitem[Avila, Rupprecht \& Beckers(1997)]{avila97}
Avila, G, Rupprecht, G., Beckers, J.
in: Optical Telescopes of Today and Tomorrow, ed. A. Ardeberg, Proc.
SPIE 2871, p. 1135

\bibitem[Andersen et al.(1995)]{andersen95}
Andersen, M.~I., Freyhammer, L. and Storm, J. 1995,
in Calibrating and Understanding HST and ESO Instruments,
ed. Benvenuti, Piero, Garching: ESO, p. 87.

\bibitem[Freudling et al.(2007a)]{freudling1} Freudling, W.,
Romaniello, M., Patat, F., M\o ller, P., Jehin, E. \& O'Brien,
K. 2007, The Future of Photometric, Spectrophotometric and
Polarimetric Standardization, ed. C. Sterken, ASP
Conf. Ser. 364, p. 113 

\bibitem[Freudling et al.(2007b)]{freudling2} Freudling, W.,
 M\o ller, P., Patat, F., et al. 2007, 
 The 2007 ESO Instrument Calibration Workshop, ed Kaufer, A. \&
 Kerber, F. (New York: Springer), p. 25

\bibitem[Freudling et al.(2007c)]{freudling3} Freudling, W.,
 M\o ller, P., Patat, F., et al. 2007, 
The Messenger, 128, p. 13

\bibitem[Fynbo et al.(1999)]{fynbo} Fynbo, J.~U., Moller, P. \&
  Warren, S.~J. 1999, MNRAS, 305, 849
\bibitem[Koch et al.(2003)]{koch03}
Koch, A., Odenkirchen, M., Grebel, E.~K., \& Caldwell, J.~A.~R. 2003
Astronomische Nachrichten Supplement, 324, 95
\bibitem[M\o ller et al.(2005)]{moller} 
M\o ller, P., J\"arvinen, A., Rupprecht, G., et al. 2005, FORS: An
assessment of obtainable photometric accuracy and outline for 
strategy for improvement, VLT-TRE-ESO-13100-3808
\bibitem[Patat et al.(2006)]{patat} Patat, F., Ugolnikov, O.~S. \&
  Postylyakov, O.~V. 2006, A\&A, 455, 395
\end{thebibliography}
\end{document}